On "Why" and "What" of randomness

Soubhik Chakraborty
Department of Applied Mathematics
Birla Institute of Technology, Mesra
Ranchi-835215
INDIA
Email: soubhikc@yahoo.co.in

If the numbers are not random, they are at least higgledy-piggledy.
- George Marsaglia (1984), quoted by D. E. Knuth on p. 179 of ref [5]

**Extended Abstract**: This paper has several objectives. **First**, it separates randomness from lawlessness and shows why even genuine randomness does not imply lawlessness. **Second**, it separates the question "why should I call a phenomenon random?" (and answers it in part one ) from the patent question "What is a random sequence?" for which the answer lies in Kolmogorov complexity ( which is explained in part two ). While answering the first question the note argues why there should be four motivating factors for calling a phenomenon random: *ontic, epistemic, pseudo and telescopic*, the first two depicting "genuine" randomness and the last two "false". The author argues that the "why" part is meant for randomness in general while the "what" part is better suited for a random sequence. Ontic in the article means *what is actual* (irrespective of our knowledge). Epistemic is *what we know (hence it relates to our knowledge of something)*. The literal meaning of pseudo is false. However the author carefully splits false into two categories: one is false *that looks like genuine* and only this is called pseudo in the article. The other is false that *looks like false* but gives us some potential benefits *if realized as genuine*. This other false randomness the author calls telescopic meaning *deliberately realized*. This new usage of the word telescopic (which normally means farseeing) needs a little understanding and practice. The words *deliberately* and *realized* are both important to understand what telescopic means in such an unconventional usage. Since telescopic randomness is a difficult concept, the author tries his best to explain it with technical and non-technical arguments. **Third**, ontic and epistemic randomness have been distinguished from ontic and epistemic probability. **Fourth**, it encourages students to be applied statisticians and advises against becoming armchair theorists but this is interestingly achieved by a straight application of telescopic randomness. Overall, it tells (the teacher) not to jump to probability without explaining randomness properly first and similarly advises the students to read (and understand) randomness minutely before taking on probability.

**Key words**: {ontic, epistemic, pseudo, telescopic} randomness; Kolmogorov complexity

PART ONE: Why should I call a phenomenon random?

**Introduction**: "*Does randomness really mean lawlessness*?" is a question I asked myself frequently and each time my conscience told me that because *we can't create an effect without a cause* (-Aristotle) there couldn't be randomness if the word were used to imply lawlessness. But we do have randomness, don't we? A careful thought explains that randomness does not quite mean lawlessness. All it means is that the experimental set up has an *inherent variability* which accounts for the *unpredictability* of the response and hence the "randomness" of the random experiment. Remember while defining a random experiment as an experiment in which the possible outcomes are known but which one will materialize cannot be predicted in advance *even if the experiment is repeated under similar conditions*, we merely said the conditions are "similar". Did we say they are *exhaustive?* Sure enough there are many more conditions in the experimental set up (that affect the response) that we did not take into account. And we did not for we have no control over them! Herein lies the "inherent variability".

A closer examination further reveals that there are two ways this "inherent variability" is caused: either the law (i.e. **the exact model used by nature**) is changing from realization to realization and thus we are unable to describe the system's behaviour by a fixed law or the law is fixed but some or all of the parameters are not. The first is precisely what the philosophers call *ontic indeterminism* or *ontic randomness*. For example, consider a complex biological process such as the response of a person complaining of a certain set of symptoms to a certain treatment on different occasions. If the law were fixed on each occasion, medical science would be an easy profession! Again it does not mean that there is no law. In every realization there is a definite law that governs the cause-effect relationship. But the law holds for that realization only and changes in the next, the next and the next! **Gentlemen, don't you feel it is this dynamic nature of the law that we are misinterpreting as "lawlessness"?**

In contrast actual coin tossing, reflecting the second case, is a purely mechanical process and therefore can be described by Newton's laws of motion (see ref. [1] and [7]), and hence it cannot be random "in principle" (that is not in the ontic sense) but is still random "in practice" (that is random in the epistemic sense) for we certainly *conceive it as random*.* Here the inherent variability is caused by the parameters some of which are clearly dynamic (parameters include the exact force applied, the point at which it is applied, the wind velocity, the parameters involved in the bombardment of the air molecules with the molecules of the coin when the coin is spinning in the air etc.……can you fix all of them in every toss?). True, Persi Diaconis, the well known mathematician and former magician, could consistently obtain ten consecutive Heads in ten tosses of a coin by carefully controlling the initial velocity and the angular momentum (ref [1])  but the question of interest is: could he do this if the law were dynamic? Moreover it is not clear whether Diaconis could achieve it if the coin after falling over the ground *rolled*.  Surely he must have tossed it over a rough surface (such as a magician's table preferably with a cloth over it). I will not enter into this debate but it is clear that Diaconis was only attempting to *fix the dynamic parameters*. The law is already fixed whence it cannot be ontic indeterminism. We call it *epistemic indeterminism in that it is not the system but the environmental factors contributing to the inherent variability in the experimental set up*. ***By environmental factors, we mean to say that there is a complex and unknown law (or even if known we are unable to write the exact model equation and solve it for a given set of parameters advance knowledge of all of which we do not have) which generates a sequence where we find hard to recognize any pattern (and hence random!), the dynamic parameters and  our  inability to smother this dynamism. But since this law is fixed it is not "in principle" contributing to the randomness howsoever complex it may be. It would have created the same response if only we could make the dynamic parameters static. It is to be understood that the inverent variability in the experimental set up is the causality and our ignorance/inabilitymentioned above a precipitating factor.***

### Classroom (ontic indeterminism versus epistemic indeterminism)

Q:  What is common to ontic and epistemic indeterminism?
 A:  In both cases the law/laws are very complex and generate sequences where it is hard to find any pattern. In both cases we don't know the exact law or are unable to write the model equation and solve it for a given set of parameters for a specific realization.

Q:  And what is the main difference?
A:  I have already said it. In ontic indeterminism, **the law is also dynamic**! In epistemic case, **the law is fixed** (e. g. Newton's laws of motion in a coin toss)

Q: You gave one example of ontic indetermisim from medical science (page 1).  Can you give another?
 A. Here is another example: the movement of an individual gas molecule is random (in the ontic sense).
Q: How do you explain it?
A: Suppose for the sake of argument, a gas molecule is at position A at one moment and goes to position B at the next and to position C at the very next. How did it come to B from A unless there is some explanation-rule? Definitely there is a law      howsoever complex it may be. The point of interest is: is the law that brought it from A to B the same as the law that brought it from B to C? If not, the indeterminism (which does not mean lawlessness) is clearly ontic. To the best of my knowledge, there is no fixed law that describes the movement of an individual gas molecule. Laws in Kinetic theory of gases are meant for a collection of gas molecules and not for an individual one!

Q:   Can I say ontic indeterminism is randomness "in principle" and epistemic indeterminism is randomness "in practice but not in principle"?
A:  Well, you can, but you have to explain that "in principle" focuses on the system's contribution and "in practice" refers to merely the environmental factors.

Q: Please differentiate between causality and precipitating factors more simply.
A: Let me give you a simpler example. Suppose you have a very bad cough and by mistake you happen to eat something that contained banana. If now you get a terrible bout of coughing, will you say it is the banana that caused it? I would say: You already have an upper respiratory tract infection (causality) that might have been be precipitated by the banana. This sounds like a better explanation. It is assumed here that you are not allergic to banana and the fruit creates no problems under normal conditions.

UPSHOT: *Genuine randomness must be caused by an inherent variability in the experimental set up in addition to complexity of the laws (**causality**), **precipitated** by our ignorance of the exact law and its parameters in specific realizations and our inability to smother the inherent variability. Since our ignorance and inability to smother this inherent variability in the experimental set up is playing a passive though definite role, it is wrong to call a phenomenon "random" out of sheer ignorance.*

**\*Remarks:-**

1. Ontology refers to the (study of) properties of a system *as they are*. Epistemology refers to the (study of) properties of a system *as we conceive*. Ontic and epistemic properties of a system can differ due to our knowledge of the system being insufficient. **We would like to make it clear that we are explaining ontic and epistemic randomness and not ontic and epistemic probability. The aim of this article is to teach the why of randomness and not probability.** However, the interested reader may refer to Hacking [11] who is often credited for coining the term "epistemic probability" meaning probability that has something to do with knowledge or information. Ontic probability finds a mention in philosophical articles and refers to the relative frequency definition of probability. Hanna [12] discusses ontic probability.

2. **We would also like to clarify here that the "why" part concerns randomness in general and not random sequences**. Thus the words "law" simply means the exact rule that explains the cause-effect relationship explicitly for the phenomenon. The reason is that words such as "law" and "parameter", which we are using in the "why" part, have different meanings when used in probability and with respect to random sequences. For example, in the context of random sequences, a fixed "law", e. g. a Poisson process with a fixed parameter, intensity λ, will produce a random sequence of outcomes. This probability model seems to contradict our views in that there is "inherent variability" without the "law" or the parameters changing! Hence a distinction between using the word "law" as applied to randomness in general as opposed to it being used for a random sequence must be made. If one extends the "why" part to a random sequence, "law" will be used to mean the exact law nature is using (static or dynamic) and "model" for the probabilistic or other approximation to this exact model the investigator is using. Thus in the coin toss, the law is Newton's law of motion and the model for a single toss is Bernoullian.

From the discussion above it follows that "genuine randomness" can be either ontic or epistemic. In fact, even when the law is dynamic, one can conceptually think of a "Master Law" (of which the changing laws are subsets) that must surely be with God but the question is: are we using the word "random" from a heavenly domain or an earthly one? The moment we use the word "unpredictability" we must clarify that "unpredictability" is *unpredictable purely from a human domain*. Agreed that the philosophical aspects of randomness are much older than the science of probability historically, but is that not a flimsy excuse the statisticians continue to provide (for escaping a philosophical discussion on randomness), given that it is they who make the maximum noise about "noise"? I write this despite being a statistician myself with all my love and respect for the community notwithstanding. If the reader is also a statistician I bring to his kind attention the sharp criticism Professor Donald Knuth (Stanford University) has made (p. 149 of ref. [5]) over this very issue. Professor Knuth has pointed out that a statement such as "the numbers behave as if they are truly random may be satisfactory for practical purposes but it sidesteps a very important philosophical and theoretical question: Precisely what do we mean by "random behavior?" A quantitative definition is needed. It is undesirable to talk about concepts that we do not really understand, especially since many apparently paradoxical statements can be made about random numbers."

Knuth continues with his characteristic style… "The mathematical theory of probability and statistics scrupulously avoids this issue. It refrains from making absolute statements, and instead expresses everything in terms of how much probability is to be attached to statements involving random sequences of events. The axioms of probability theory are set up so that abstract probabilities can be computed readily but nothing is said about what probability really signifies, or how this concept can be applied meaningfully to the actual world. In the book Probability, Statistics, and Truth (New York: Macmillan, 1957), R. von Mises discusses this situation in detail, and presents the view that *a proper definition of probability depends on obtaining a proper definition of a random sequence*." (see also the author's final remark at the end of this article)

Having covered "genuine randomness" I now come to that randomness which is "false". Interestingly, even false randomness can be of two types. The first is the type that is false but *looks like genuine*. That is to say, we have a *deterministic known model* working in every trial but the model is so developed that the sequence of the response has some statistical properties of randomness similar to that which we expect in an ontic or epistemic case (like the number of "runs" being neither too small nor too large) where the model may not be in our hand. This is the pseudo random case (pseudo means false) which is at best only statistically random; two well known examples which I demonstrated in the (post-graduate) class include the (pseudo) random numbers generated by the computer by Lehmer's linear congruential method and by Tausworthe's shift register method. See ref. [4] for details.

**Classroom Exercises:** For the linear congruential model $r_{i+1} = (ar_i + b) \mod m$ (where $m>0$, $0<=a<m$, $0<=b<m$, $0<= r_0 <m$) I gave students exercises to verify, by selecting numerical values of a, b and m, that

(i) within a "cycle" the sequence does look like random and *therefore these can be used to simulate ontic or epistemic randomness provided only the entire simulation work is over within the cycle* [herein lies another philosophy: it makes sense to create randomness with a known model than an unknown one (-see Knuth)] and therefore that we must vow not to waste random numbers. Once a cycle is complete, the same sequence follows.

(ii) an arbitrarily selected seed $r_0$ may not give random numbers. I advised the students to allow the computer itself to select the seed as a detailed discussion on this might destroy interest.

(iii) The constants a (called a priori) and b (called multiplier) must be selected to maximize the "period" which at most can be m (called modulus which is some large positive integer) as it comes from the remainders 0, 1, 2…m-1 in a desirable statistically random order (which are then divided by m to give an approximate U[0, 1] distribution).

The students completed the exercises with little difficulty. For example it was nice to see one of them observing that a=b=1 always maximized the period but the sequence is never random! I pointed out a theorem (suppressing the proof) which asks b to be relatively prime to m, a-1 to be a multiple of p for any prime factor p of m and a-1 to be a multiple of 4 if m is a multiple of 4 (for proof, see Knuth [5])

One student enquired whether it is possible to have sequence where there is no cycle. I said it is certainly possible pointing out that the digits in the decimal expansion of irrational numbers have this property. I further told him that pi has already been proposed as a random number generator and the allegation that it is "less random than we thought" has been nullified by George Marsaglia (see [14] and the reference cited therein).

Remark: The (pseudo) random generator defined here is called multiplicative if b=0 and mixed otherwise.

The fourth and last category of randomness is the second category of false randomness which is *neither random nor does it look like it is random*. However if it is *realized* as random we get some potential benefits such as *the prediction becoming less costly and more efficient*. There are situations in which we do look for a cheaper and more efficient predictor e.g. in a computer experiment, which is a series of runs of a code for various inputs. I permit myself to call this last category *deliberately realized indeterminism* or *telescopic indeterminism*. Just as a telescope reduces the cost (physical effort, time, money etc.) of actually reaching a distant object and *we agree to realize that it is coming near (though it is not!) by looking through the telescope*, similarly *by deliberately realizing the non-random data as the outcome of a stochastic process we can reduce the cost of prediction and make prediction more efficient in a computer experiment*. (ref[2]). **It is important to note that the experimental side of algorithmic complexity is only a special kind of a computer experiment in which the response variable is a complexity (some resource consumed, while executing the code, such as time).** While deterministic prediction requires knowledge of the entire input, for a stochastic prediction it suffices to know the size of the input which amounts to knowing some input parameter(s) that characterizes the size. This ability of the statistician to predict with partial knowledge of the input explains *efficient prediction* (especially in complexity analysis of algorithms where algorithmic complexity itself is expressed only as a function of input parameter characterizing the input size). *Cheap prediction* here simply means the ability of the statistician to predict the response for huge untried inputs/input size for which it is computationally cumbersome to both feed the code and run it for such inputs. Designing a computer experiment amounts to efficient choice of input sites. For more on the link between algorithmic complexity and computer experiments, see ref [3].

**Distinguishing telescopic randomness from non-telescopic randomness through classroom exercises:-**

Fortunately several of the exercises I gave in the class on this intriguing topic are already available in published papers.

### Classroom

I asked one group of postgraduate students to write programs on Replacement sort and another group on Winograd's algorithm on n x n matrix multiplication respectively and measure the number of interchanges for the first and running times for the second at different points of n, for randomly varying input elements taking several readings at each point of n. They were asked to report their observations to me. They gave me the following reports :-

First Group (who programmed Replacement sort in C and BASIC): It is easy to generate a good quality of "noise" at each point of n. It makes sense to characterize the response by a stochastic model as a function of n.

I explained: Yes, and this is because in a sorting algorithm *fixing n does not fix the computing operations* which also depend on the input elements and their relative positions. Hence the deterministic response which comes for a fixed input may be taken as stochastic for a fixed n and randomly varying input elements. The argument is definitely true for algorithms such as sorting and searching. I pointed out to the students that this is the argument Professor Hosam Mahmoud (George Washington University) has given in the preface of his book in ref. [9]. In ref. [3] I have called it *the traditional school of thought* (for defending stochastic modeling). In this case fixing n fixed the comparisons = $n(n-1)/2$ since the simplified version of replacement sort [3] was used but the interchanges were certainly not fixed. They generated the desired "noise" for each n.

Second Group (who programmed Winograd's algorithm): Even for higher n (which means more input elements) it is hard to generate "noise" for repeated trials at each n. Most readings are identical and even for those which are not, the negligible difference makes it difficult to ascertain as to whether it is caused by the idiosyncrasies of computer clocks or a genuine random behavior of the time response. The students enquired whether a stochastic model can be fitted here also and, if yes, on what ground?

I explained: In Winograd's algorithm on nxn matrix multiplication, fixing n *fixes all the computing operations*. That is why it hard to generate "noise" at each point of n simply by manipulating the operands (the matrix elements).*The time consumed by an operation depends on three factors: the operation type, the system and the operands.* But only the first two are major contributors. I am not saying the operands do not contribute to time. They do. For example, operands consuming fewer bytes will require less time for their storage in the memory. But such contributions are not enough to create "noise" in the sense of randomness or at least statistical randomness! Regarding fitting a stochastic model, you can certainly fit but the argument for defending it has to be different from that in the sorting algorithm case. **All you need to say is that as your experiment is a computer experiment so you are only looking for a cheaper predictor… cheaper and efficient...and proceed.** I pointed out to the students that this is the same argument Professor Jerome Sacks (now a Professor Emeritus at Duke University; formerly Professor and Head, Department of Statistics, University of Illinois) gave in a landmark paper [2] which prominent statisticians such as Michael Stein "wholeheartedly" supported criticizing traditional statisticians, accustomed to fitting stochastic models only to random data, for shying away from untraditional research (fitting stochastic models to non-random data). (I too seriously used the word "untraditional" in [3] and no pun is intended)

UPSHOT: The randomness in the second case is *telescopic* since it is neither random nor does it look like. In the first it is at least statistically random We want a model with n only as predictor and expect some statistical properties of randomness from such a model and make use of them. If we wanted exact prediction, which corresponds to a specific input, a deterministic model is to be sought.

Q: What is the main difference between pseudo and telescopic randomness?
A: **The difference between telescopic randomness and pseudo randomness is that while both are false, pseudo randomness at least looks random and therefore it can be used to simulate genuine randomness. Telescopic randomness does not even look random! Imagine a phenomenon generating observations as say …..…40, 40, 40, 40, 40……**where it is understood that all repetitions on the left and right are 40 as well. While an armchair theorist would say "Sure Statistics, which is a study of variation pertaining to numerical data, has nothing to do such phenomenon. Where is the variation?" an applied statistician (such as Sacks and Stein) has other ideas. Most computer experiments are deterministic and this is true regardless of whether the response is the output or a complexity. Then why are they characterized by stochastic models? I have already answered this question earlier and would simply now refer to [2] and [3] for more on it.

BENEFITS: *The first paper in any peer reviewed journal which connects algorithmic complexity with computer experiments has two of my post graduate students partnering me (see ref. [3]).* The first such paper in any journal is possibly the popular science article [10] where again my young friends have partnered me. Winograd's algorithm results are also published [8]. I encouraged the students to seek other situations where telescopic indeterminism might apply.

As a final comment, this note answers the question: *Why should I call a phenomenon random (even if it is not!)?* Observe that this is different from the patent question: *What is a random sequence and what is not?* [5] The first question seeks the possible motivations for calling **a phenomenon** random (and I have argued there can be four). The second demands a strict qualification for **a random sequence** and seeks a mathematical tool for the purpose. For the latter question the straightforward answer is: **A sequence is random if the length of the shortest program which outputs the sequence (technically called the *Kolmogorov complexity* of the sequence) equals the length of the sequence.** But this is an established concept (thanks to Knuth again among others; See also [6]) and there is hardly any room for fresh arguments except write a non-technical note on it which is easily grasped compared to Knuth's "bible"! This is in fact what I promise to do in part two of this article. The present article is meant for statistics students **exclusively** whereas Kolmogorov complexity is more about computing science and applied mathematics than statistics and probability. I therefore sensibly concentrated on the "why?" part only here which relates to a phenomenon being classified as random (either in true or false sense) on the basis of motivating factor rather than a specific random sequence for which the "what" part is ideal. I shall however explain later why the "what" part is more meaningful for a random sequence than for general randomness.

I also attempted to teach telescopic randomness at the undergraduate level in a non-technical way citing other situations where we opt for realizations rather than reality like cinema, poetry and even war and linking them with statistics. For example, the students found it quite amusing when I said "Think of the actor Ben Kingsley *realizing* himself as Mahatma Gandhi in Richard Attenborough's classic movie "Gandhi", think of Nepoleon asking his soldiers "Where is Alps? I don't see any Alps! Go ahead!". Clearly *there was Alps* blocking the soldiers. This is reality. What Nepoleon asked his soldiers to do is a *realization*. Similarly Ben Kingsley is clearly not the Mahatma but let us face the truth; the original Mahatma is no more and *the tale has to be told by a living man*. And that is precisely what Kingsley did while acting under the jurisdiction of the director. The money a professional actor earns, the pleasure and personal satisfaction he derives from acting and the healthy entertainment he provides in return are the potential benefits. Think of the anonymous Hindi poet who breaks the rhythm to write "*Tum mujhe kya kharidoge? Main muft hoon!*" (translated free, it reads: *In what way will you buy me?/ I am free*!). I ask: do poems sell because of rhythm or do they sell because of ideas? If it is the ideas that are more important in the final analysis, why should a stochastic model depend exclusively on the randomness of the data? Why not on the statistician's motivation (idea) which can be justified in contexts even where there is a total absence of randomness? (this link between poetry and stochastic modelling is available in [3])

The periphery of reality is quite limited in that if there is water in the glass, there is water and if there is no water then there is no water. This is reality. Imagination has a much wider periphery. Take an empty glass in summer, imagine there is water and pretend to drink it. Does it lift your spirits? I have tried this. It does! (if it does not, imagine wine in place of water!). The point is: statistics *can be made* and *I define an applied statistician not as one who merely applies statistics in other areas (for which the nomenclature "statistical practitioner" suffices) but as one who wholeheartedly believes in **making statistics even where there is none** rather than wait for nature to provide a tailor-made situation for him*. Note 1: The words *deliberately* and *realized* are both important to understand what telescopic means in such an unconventional usage. Since telescopic randomness is a difficult concept, I tried my best to explain it with technical and non-technical arguments. Nevertheless, the beginner is advised to use the nomenclature *deliberately realized indeterminism* initially and replace it by its synonym telescopic indeterminism only after gaining some confidence.

Note2: Strictly speaking, cinema is both pseudo and telescopic. It is pseudo in that Kingsley dressed and behaved as Gandhi and therefore *looked like the original Gandhi* on screen. It is telescopic in that he had to *realize* himself as Gandhi and not Kinsley in order to behave like Gandhi to some perfection. It is to be understood that the realization part is more important. You or I can dress like Gandhi, cut all the hair, wear spectacles, walk with a stick etc. But how many of us can act like Kingsley?

**Final remark**: Most teachers and authors of probability books begin with a shallow explanation of randomness and jump to probability thinking it is possible to explain probability meaningfully even with a workable explanation of randomness. It is not. Probability can be meaningfully appreciated provided only randomness is properly understood first. Further, a proper knowledge of randomness is never acquired without learning both the "why" and the "what" part. The "why" part is a fairly general question and therefore more suited for a phenomenon rather than a specific sequence in hand. But it is more challenging to ask the "what" part for a sequence rather than a phenomenon. For, if the question is "What is a random phenomenon in the true sense?" the simple answer would be "One which is random either in the ontic or the epistemic sense". Similarly to the question "What is a random phenomenon in a false sense?" the simple answer is "One which is random either in a pseudo or a telescopic sense". But for a random sequence the "what" part assumes a special significance and has to be dealt with separately because *the galaxy of statistical tests on randomness (gap test, poker test, run test to name only a few) taken collectively fail to certify randomness of a sequence in the strict sense and merely certify that the sequence "looks like random"*. Although it serves some purpose (e.g. simulating genuine randomness), as mentioned earlier, such 'looking like' activities are disliked by celebrated mathematicians and I therefore thank Professor Knuth among others [6] for painstakingly explaining the "what" part of a random sequence in ref. [5] while I took on the more general "why" part here to complete the "noisy" story. As mentioned earlier, the "what" part of a random sequence has more to do with computer science and mathematics rather than statistics as here the basic question **"what has a computer program to do with a random sequence?"** has to be addressed to the students. I shall now provide a non-technical note on it in part two.

I end this part with two memorable quotes:

**God does not play dice.**
          **-----Albert Einstein (Creator and Rebel, 1973, p. 193)**

**God not only plays dice, He also sometimes throws the dice where they cannot be seen.**
          **----- Stephen W. Hawking (Nature, 1975, 257, p. 362)**

And I leave it to the reader to resolve the contradiction between the two quotes and hence discover the contribution which I believe he should be able to do if he has understood this part. [Part one Concluded]

## PART TWO: What is a random sequence?

It was stated (in part one) that a sequence is random in the strict sense if the length of the shortest program which outputs the sequence (called the Kolmogorov complexity of the sequence) equals the length of the sequence itself. Part two of this paper explains the meaning of this definition in a friendly tone. It answers the fundamental question *"What has a computer program to do with a random sequence?"*. To understand the definition we must clarify why we are interested in the shortest program which outputs the sequence and not any arbitrary program and what role the length of the program plays in determining randomness.

Suppose a BASIC programmer writes
PRINT "AB";
END
(this will run perfectly in QBASIC environment with AB as the output; statements need not be numbered in QBASIC) It looks he has written a small program. A moment's reflection however suggests that the length of the program is somewhat greater than the length of the sequence for the entire sequence has been written inside the program! And what if the sequence had 1000 characters? Would he write PRINT "ABABAB….AB"( each letter repeated 500 times)?
Using intelligent programming we could certainly do better as follows:

FOR i=1 TO 500
PRINT "AB";
NEXT i
END

Only 31 characters are used in the code above to output the desired sequence!
If the sequence ABAB…had 10000 characters we would simply replace 500 in the code by 5000. In general, to generate AB… n times, n itself will have to be specified by some positive integer. If w is the largest positive integer such that $10^w \leq n$, i.e., $w \leq \log_{10} n$ then $\log_{10} n$ is the approximate number of characters required to specify n which means the length of the code above in the general case is roughly $28 + \log_{10} n$ which for large n is small in comparison to n.

### Classroom

Student: It seems we cannot create a program, which outputs a sequence, shorter than the number of characters needed to specify the sequence.

Teacher: Correct. In the example above we only tried to get close to that figure. We could easily write longer codes which would output the same sequence, e.g., using some empty FOR…NEXT loops and therefore it makes sense to consider the shortest programs only and not all programs for explaining randomness. I hope you have understood this.

Student: Indeed I have but there is another important lesson I have learnt from this example. Because the sequence ABAB… had a definite pattern so we could write a short program to generate it a large number of times using a programming strategy (counter controlled looping technique). It seems logical to argue that a sequence with no pattern would require a larger program for its generation than a sequence with patterns. This explains why the length of the program is taken into consideration (rather than the underlying algorithm's complexity such as time or space) for defining randomness. But is there a sequence which cannot be generated by any program (howsoever it is written) except that whose length is approximately the length of the sequence itself? That would be a "truly random" sequence!

Teacher: Yes, there can be such a sequence. But before we prove the existence of such a sequence, we have to first clarify what programming language is being used. BASIC programs are shorter than those in COBOL or FORTRAN generally. Again, celebrated computer scientists such as Professor Donald Knuth (Stanford university) prefer to work on machine language only (0 and 1). Since all programs, whatever be the programming language used, are ultimately converted to 0 and 1 only, therefore there is no harm in confining ourselves to machine code only. Further we make a rule that no program can be a prefix of another e.g., both 101 and 101110 cannot be valid as 101 is a prefix in 101110.

Now let us consider all possible sequences of 0's and 1's of fixed length n. There can be $2^n$ such sequences. Since each program can output only one sequence and no program can be a prefix of another, there are at most $2^r$ sequences of length r whence the number of programs of length less than n is at most $2^0 + 2^1 + 2^2 + \ldots + 2^{n-1} = 2^n - 1$.

As each program is outputting only one sequence, as mentioned earlier, from the above we conclude that there is at least one sequence of length n which is not outputted by the programs of length less than n, that is to say, which will require a program of length n or higher. This completes the proof.

Remark: As $\frac{1}{2} \times 2^n = 2^{n-1}$ and $2^0 + 2^1 + 2^2 + \ldots + 2^{n-2} = 2^{n-1} - 1$ it follows that half the sequences of length n will require programs of length n-1 or higher for their generation.
This means quite a good number of sequences are almost random!

Classroom

Student: I have understood now what a computer program has to do with a sequence being random and that there are random sequences. But can you give me an example of a random sequence? It has to be "random" in the sense of Kolmogorov complexity.

Teacher: In fact this discussion will be incomplete without such an illustration. The desired random sequence will be an infinite sequence of 0's and 1's in the programming language that we agreed upon where no program can be a prefix of another. To generate the first n bits of this sequence, we will require a program of length greater than n-k. We shall see what value k takes. You will find it to be like the constant 28 in the BASIC code discussed earlier.

Let us group the programs into two: those that halt (after generating a sequence) and those that do not, i.e. they enter some infinite loop. Next consider the totality of all programs from both the groups. Let us inspect each program and if it halts, compute $1/2^u$
where u is the program's length. Our desired sequence is the binary equivalent of the sum* of all such terms of the form $1/2^u$ after deleting the (binary) point.

(*this sum is a number between 0 and 1 by Kraft's inequality. We omit the proof.)

Student: But how do we verify that this is the desired random sequence?
Teacher: Let us call this sequence N. To verify that N is indeed random, we have to utilize the fact that the N contains information as to which programs halt and which do not. Observe that N is the sum of terms of the form $x/2^y$ where x is the number of programs of length y that halt. Since no program can be a prefix of another it means that the numbers for the different lengths do not get jumbled up in constituting N and each can be successfully extracted!

Now suppose there is a program P which generates the first n bits of N. Further let Q be a program which has P as a subroutine and the main routine has k characters. N may be taken as a random sequence if we can show that the length of P must be greater than n-k.
Student: That sounds interesting. But how will you prove it?
Teacher: The main routine first calls P and gets the first n bits of N. Using the information contained in the first n bits of N, the main routine figures out how many programs (of length at most n) halt. Suppose this number is T. It is also able to figure out how many programs (of length at most n) do not halt. This is done by picking up each program and executing it until it halts with a counter being kept that keeps increasing by one each time a halting program is discovered. The final value of the counter reads T. Next, for each program that halts, the main routine determines the sequence outputted by this program. Finally the main routine determines a sequence which is different from all the sequences outputted by halting programs of length at most n. This sequence must have Kolmogorov complexity more than n in that it is not generated by any of the programs of length at most n that halt. Evidently, we have length of Q = length of P + k. Since the final output of Q is a sequence with Kolmogorov complexity more than n, the length of Q must be more than n = n + f (say) whence n + f = length of P + k which means length of P = n-k + f > n-k.
Q. E. D.!

Student: I have understood the proof alright but I still have some queries. First, suppose the main routine picks up a program that does not halt by mistake. It will get stuck!

Teacher: Yes, it will! To prevent this, the main routine will first execute each program *for one step*. Then each program *which has not yet halted* is run for two steps. Then each unhalted program is similarly run for three steps and so on. The process continues until T programs have halted. The main routine now knows which programs have halted and hence the remaining must be those that do not halt. [13]

> **Remark**: There is a famous halting problem (proved by Turing) which states that there is no program which can check whether another program halts. If such a program existed, our task would be a cake walk. We could easily have a short program P to compute the first n bits of N, for each n, by first counting all programs of length n or less, then deciding for each such program as to whether it halts or not and sum terms of the form $1/2^u$ to N for each program of length u that halts. Observe how our inability to do so has been combated in the discussion above. [Concluded]

**Acknowlegement**: This article is dedicated to the untraditional revolutionist Prof. Jerome Sacks and the inimitable Prof. Donald Knuth. I thank Prof. William Notz and an anonymous referee for several suggestions. I thank all the authors wholeheartedly whose works I have freely consulted.